\documentclass[a4paper,12pt]{article}
\usepackage[utf8]{inputenc}
\usepackage{cancel}
\usepackage{ulem}
\usepackage{amsfonts}
\usepackage{amssymb}
\usepackage{graphicx}
\usepackage{amsmath}
\usepackage{enumerate}
\usepackage{mathtools}
\usepackage{subfig}
\usepackage{color}
\usepackage{tikz}\usetikzlibrary{calc}
\usepackage{float}
\usepackage{here}
\usepackage{cite}
\usepackage{mathrsfs}
\usepackage{float,epsfig}
\usepackage{dcolumn}% Align table columns on decimal point
\usepackage[toc,page]{appendix}

\usepackage{graphicx}% Include figure files
\usepackage{bm}% bold math
\usepackage{amsmath,amssymb,amsthm}
\usepackage[colorlinks=true,linkcolor=blue,citecolor=red]{hyperref}
\newcommand{\be}{\begin{equation}}
\newcommand{\ee}{\end{equation}}
\newcommand{\bea}{\setlength\arraycolsep{2pt} \begin{eqnarray}}
\newcommand{\eea}{\end{eqnarray}}

\def\0{{\sst{(0)}}}
\def\1{{\sst{(1)}}}
\def\2{{\sst{(2)}}}
\def\3{{\sst{(3)}}}
\def\4{{\sst{(4)}}}
\def\5{{\sst{(5)}}}
\def\6{{\sst{(6)}}}
\def\7{{\sst{(7)}}}
\def\8{{\sst{(8)}}}
\def\sst#1{{\scriptscriptstyle #1}}

\makeatletter \@addtoreset{equation}{section}

\usepackage{multirow}
\definecolor{lime}{HTML}{A6CE39}
\tikzset{>=latex}

\begin{document}
\title{{\normalsize \textbf{\Large   Constraining Black Hole Shadows in Dunkl Spacetime  using  CUDA  Numerical  Computations 
 }
}}
\author{ \small  Saad Eddine Baddis,     Adil  Belhaj, Hajar Belmahi \footnote{hajar\_belmahi@um5.ac.ma}, 	and   Maryem  Jemri \thanks{
Authors are listed  in alphabetical order.} \hspace*{-8pt} \\
%EndAName
{\small ESMaR, Faculty of Science, Mohammed V University in Rabat, Rabat, Morocco  } }
\maketitle

\begin{abstract}
With the help of   CUDA high-performance numerical codes exploited in machine learning,   we investigate   the shadow aspect of  new  rotating and   charged black holes using  the Dunkl derivative formalism.     Precisely, we  first  establish  the corresponding metric function encoding  the  involved physical properties  including the  optical character.   Exploiting such accelerated simulations, we  approach     the horizon radius behaviors   in order to determine the   regions of the module space providing physical solutions.    Applying the Hamilton-Jacobi mechanism,  we assess      the shadow  aspect   for  non-rotating and rotating solutions. Using such an aspect,  we evaluate    the   energy  rate of emission.  Developing   a high-performance CUDA numerical code, we derive strict constraints on  the Dunkl deformation parameters in order to establish a link with the shadow observations provided by the Event Horizon Telescope collaboration.
 {\noindent} 

\end{abstract}

%	\begin{flushright}
%	ksdjflk kjsdlfkj
%	\end{flushright}
%

\newpage

\section{Introduction}
Black holes in general relativity  are considered  as solutions of  the Einstein  field equations  being  one of the most fascinating predictions of modern physics and related topics \cite{1}. Beyond being gravitational objects,  there has been a lot of effort  to  focus on their thermodynamical properties, like phase transitions and critical phenomena.  In addition,   their optical  behaviors  governing the light  motion   in their strong gravity field have  been  largely  investigated using analytical and numerical approches furnishing  promoting  predictions  which could be tested via falsification scenarios \cite{2,80,81,82,83,84,85,86}. To this purpose, the Event Horizon Telescope (EHT) has realized record-breaking measurements by imaging the shadow of the supermassive black hole, connecting the astrophysical observations and the  theoretical models \cite{3,4,5,6,7}.  Alternatively,  the study of the  black hole shadows has been developed not only in ordinary spacetimes but also in deformed spacetime geometries including the implementations of non-trivial couplings   like the Gauss Bonnet  scenarios\cite{700,701,702,703}.

Recently,  certain black holes have  been studied in situations of spacetime  modifications. Such  deformations   could  be caused by  geometric  fluctuations, noncommutative spaces or Dunkl-type differential operators  \cite{8,9,10,11}. Alternatively,  certain deformations have been generated by  implementing   certain physical fields including  dark energy and dark matter \cite{12,13,14,15,16,17,18}. Roughly,  the deformation scenarios allow the  spacetime structure  to be modified.  In this way, the black hole shadows  and the  photon trajectories  their vicinity  can  be altered. The deformations induced by  physical  fields, on the other hand, indirectly affect the parameters of the black hole by reworking  the energy profile in its vicinity \cite{19}.

In this regard, extremely charged black holes, or more precisely the Reissner-Nordström solution and its deformed equivalents, have been the subject of extensive studies  in recent years\cite{20}. Their shadows have  been studied, and  the corresponding  investigation has highlighted that the electromagnetic charge leaves a recognizable trace. More specifically, it  makes the shadow more asymmetrical compared to that of the neutral Schwarzschild black hole of equal mass \cite{21}. This offers a potential observation technique to deduce the charge-to-mass ratio of  the astrophysical black holes, even though their total charge is supposed to be zero in standard scenarios. On the contrary, the investigation proceeds into the curved spacetime, where the charge is linked   to  the geometric curvature or  the external fields\cite{22}. In the lack of a cosmological constant, modifications to the shadow of a charged black hole may still arise from other theoretical frameworks, such as an enveloping quintessence field or noncommutative geometry. In these cases, the shadow is significantly affected. The radius may constrict or expand, and its shape may exhibit distinctive distortions. These are significant analyses as they establish a subtle correspondence between the  shadow observables and the internal parameters of the black holes. They thus provide a powerful multi-parameter testing framework  to  explore exotic physics including    general relativity extensions  \cite{23}.

The aim of this work is to investigate  the shadow of   new   rotating and charged black holes by combining  the  Dunkl derivative formalism  and    CUDA high-performance numerical codes being  explored in machine learning.    To do so, we  first  establish the corresponding metric function encoding the physical properties  including the optical  behaviors.  Using the CUDA-accelerated simulations, we approach the behaviors of the horizon radius  in order to determine the   regions of the module space providing physical solutions.   With the  help  of  the Hamilton-Jacobi mechanism,  we investigate the shadow of  non-rotating and rotating solutions. Then, we  calculate  and  examine   its energy rate of emission. In order to enable a  direct comparison with current astrophysical data, we exploit a CUDA high-performance numerical code to generate constraints on shadows.  Precisely, we  provide a   numerical approach  which enables a systematic study  by imposing strict limits on the Dunkl deformation parameters   matching with   the shadow observations provided  by the EHT collaboration. 

The organisation of the paper is the following. In Section 2, we  first establish  a  theoretical framework to derive the charged black hole metric using  the Dunkl deformed spacetime formalism.  Then,  we approach  such behaviors  using CUDA techniques. Section 3 concerns  the  shadow of non-rotating and rotating charged   black holes. In Section 4, we numerically estimate the energy emission rate  needed to analyze its Hawking radiation behavior. In Section 5, we assess the   current theoretical predictions by linking them to EHT observations using CUDA-accelerated simulations.  In the last section, we provide  some concluding remarks.
\section{ Rotating and charged black holes in the Dunkl spacetime}

In this section, we  elaborate      charged black hole solutions in a  Dunkl spacetime by considering the associated extended derivatives relying on  reflection symmetries. Concretely,  we deal with a charged spacetime by combining  certain Dunkl deformations,  the gravitational,  and  the electromagnetic fields.

 Considering a static, spheric,  and symmetric spacetime,   the charged Dunkl   black hole is assumed to be described  by the following metric  function form 
\begin{equation}
\mathrm{d}s^2 = g_{\mu\nu} \mathrm{d}x^{\mu} \mathrm{d}x^{\nu}=-f(r)\,\mathrm{d}t^2 + \frac{1}{f(r)}\,\mathrm{d}r^2 + r^2\left(\mathrm{d}\theta^2 + \sin^2\theta \,\mathrm{d}\phi^2\right),
\end{equation}
where \( f(r) \) is an unknown function which can be obtained   by solving the Einstein field equations in the Dunkl spacetime.  The latter refers to a generalized spacetime model resulting from the application of the  Dunkl derived formalism. This modifies  the standard  derivatives by incorporating reflection symmetries. Usually, the  function  \( f(r) \) depends on the internal and  the external   black hole parameters leading to  solutions going beyond the ordinary ones.   A priori, there are  many roads to approach  such a function  via certain modifications supported either by the spacetime geometry or the  involved physical matter. Among others,  many geometric  deformations have been proposed   providing certain extended metric functions.  Here, roughly,   we reconsider the study  of  the modified geometry induced by Dunkl operators  which can be expressed as follows
\begin{eqnarray}\label{4}
	D_{x_{\mu}}= D_{{\mu}}=\dfrac{\partial}{\partial x_{\mu}}+\dfrac{\alpha_{\mu}}{x_{\mu}}(1-\mathcal{R}_{\mu}),\quad\quad \mu=0,1,2,3
	\end{eqnarray}
where  $\alpha_{\mu}$ and $\mathcal{R}_{\mu}$ denote the Dunkl parameters and the parity operators, respectively \cite{25,26,27,28}.   Such a  Dunkl derivative formalism  is a generalization of the ordinary  one  in the Euclidean spaces involving
r­eflection symmetries. These extended  derivatives introduced by C. F.
 Dunkl  are based on   re­flection operations  corresponding to  non-trivial  algebraic structures  such as  the Coxeter groups and the root systems exploited in the classification of Lie algebras \cite{2808,2809}.

 In order to establish  the associated equations of motion in the proposed black hole physics,  it  is convenient  
to employ  the spherical coordinates.   Considering   $\alpha_{\mu} = \left(0, \alpha_{1}, \alpha_{2}, \alpha_{3}\right)$ and $\mathcal{R}_{\mu} = \left(0, \mathcal{R}_{1}, \mathcal{R}_{2}, \mathcal{R}_{3}\right)$,  the Dunkl operators   can take the following form 
	\begin{equation}\label{6}
	\begin{aligned}
	&D_{r}=\frac{\partial}{\partial r}+\frac{1}{r}\sum_{i=1}^{3}\alpha_{i}(1-\mathcal{R}_{i}),\quad\quad D_{t}=\frac{\partial}{\partial t},\\
	&D_{\theta}=\frac{\partial}{\partial \theta}+\sum_{i=1}^{2}\alpha_{i}(1-\mathcal{R}_{i}) \cot\theta-\alpha_{3}(1+\mathcal{R}_{3})\tan\theta,\\
	&D_{ \phi}=\frac{\partial}{\partial \phi}-\alpha_{1} \tan\phi (1-\mathcal{R}_{1})+\alpha_{2} \cot\phi (1-\mathcal{R}_{2}).
	\end{aligned}
	\end{equation}
To get the metric function $f(r)$, certain geometric quantities should be determined.  Indeed, one needs to   calculate  the Christoffel symbols given by 
\begin{equation}
\Gamma^\lambda_{\mu\nu} = \frac{1}{2} g^{\lambda\rho} \left( D_\mu g_{\nu\rho} + D_\nu g_{\mu\rho} - D_\rho g_{\mu\nu} \right).
\end{equation}
After calculations, the non-zero modified Christoffel symbols are   found to be 
{\footnotesize
\begin{equation}
\begin{aligned}
\Gamma^{r}_{rr} &= \frac{1}{2} \left( -\frac{f'}{f} + \frac{1}{r} \sum_{i=1}^{3} \alpha_i (1 - \mathcal{R}_i) \right), \\
\Gamma^{r}_{\theta\theta} &= -\frac{f r}{2} \left( 2 + \sum_{i=1}^{3} \alpha_i (1 - \mathcal{R}_i) \right), \\
\Gamma^{\phi}_{\theta\phi} &= \Gamma^{\phi}_{\phi\theta} = -\frac{1}{2} \left( 2 \cot\theta + \sigma \right), \\
\Gamma^{r}_{tt} &= \frac{f}{2} \left( f' + \frac{f}{r} \sum_{i=1}^{3} \alpha_i (1 - \mathcal{R}_i) \right), \\
\Gamma^{\theta}_{r\theta} &= \Gamma^{\theta}_{\theta r} = \frac{1}{2r} \left( 2 + \sum_{i=1}^{3} \alpha_i (1 - \mathcal{R}_i) \right), \\
\Gamma^{\theta}_{\phi\phi} &= -\frac{1}{2} \sin^2\theta \left( 2 \cot\theta + \sigma \right), \\
\Gamma^{t}_{tr} &= \frac{1}{2} \left( \frac{f'}{f} + \frac{1}{r} \sum_{i=1}^{3} \alpha_i (1 - \mathcal{R}_i) \right), \\
\Gamma^{\phi}_{r\phi} &= \Gamma^{\phi}_{\phi r} = \frac{1}{2r} \left( 2 + \sum_{i=1}^{3} \alpha_i (1 - \mathcal{R}_i) \right), \\
\Gamma^{r}_{\phi\phi} &= -\frac{f r \sin^2\theta}{2} \left( 2 + \sum_{i=1}^{3} \alpha_i (1 - \mathcal{R}_i) \right), 
\end{aligned}
\end{equation}
}
where  $ \sigma$  is  a parameter written as 
\begin{equation}
\sigma= \sum_{i=1}^{2} \alpha_i (1 - \mathcal{R}_i) \cot\theta - \alpha_3 (1 + \mathcal{R}_3) \tan\theta
\end{equation}
carrying    data on the   spherical Dunkl deformations. 
Exploiting such modified Christoffel symbols, we can calculate the Ricci tensor \begin{equation}
R_{\mu\nu} = D_\alpha \Gamma^{\alpha}_{\mu\nu} - D_\nu \Gamma^{\alpha}_{\mu\alpha} + \Gamma^{\alpha}_{\mu\nu} \Gamma^{\beta}_{\alpha\beta} - \Gamma^{\alpha}_{\mu\beta} \Gamma^{\beta}_{\alpha\nu}.
\end{equation}
The non-zero components  are shown  to be 
\begin{equation}
\begin{aligned}
R_{tt} &= \frac{1}{2} f f'' + \frac{f f'}{r} \left( 1 + \frac{3}{2} \sum_{i=1}^{3} \alpha_i (1 - \mathcal{R}_i) \right) + \frac{f^2}{r^2} \sum_{i=1}^{3} \alpha_i (1 - \mathcal{R}_i) \left( \frac{1}{2} + \sum_{i=1}^{3} \alpha_i (1 - \mathcal{R}_i) \right), \\
R_{rr} &= -\frac{1}{2} \frac{f''}{f} - \frac{f'}{f r} \left( 1 + \frac{3}{2} \sum_{i=1}^{3} \alpha_i (1 - \mathcal{R}_i) \right) - \frac{3}{2 r^2} \sum_{i=1}^{3} \alpha_i (1 - \mathcal{R}_i) \left( 1 + \sum_{i=1}^{3} \alpha_i (1 - \mathcal{R}_i) \right), \\
R_{\theta\theta} &= -f - r f' + 1 - \frac{r f'}{2} \sum_{i=1}^{3} \alpha_i (1 - \mathcal{R}_i) - \frac{5}{2} f \sum_{i=1}^{3} \alpha_i (1 - \mathcal{R}_i) - f \left( \sum_{i=1}^{3} \alpha_i (1 - \mathcal{R}_i) \right)^2 + \delta, \\
R_{\phi\phi} &= R_{\theta\theta} \sin^2\theta,
\end{aligned}
\end{equation}
where $\delta$ is  a corrected Dunkl  parameter  given by 
\begin{equation}
\delta =\sigma \left( \frac{\sigma}{2} + 2 \cot\theta \right).
\end{equation}
Contracting the Ricci tensor with the inverse metric,  we get   the  Ricci scalar
\begin{equation}
R =-f'' - \frac{2f}{r^2} + \frac{2(1 + \delta)}{r^2} - \frac{4f'}{r} \left( 1 + \sum_{i=1}^{3} \alpha_i (1 - \mathcal{R}_i) \right)
- \frac{f}{r^2} \sum_{i=1}^{3} \alpha_i (1 - \mathcal{R}_i) \left( 7 + \frac{9}{2} \sum_{i=1}^{3} \alpha_i (1 - \mathcal{R}_i) \right).
\end{equation}
To obtain the black hole   charged solution that we are after,   we consider the Einstein-Maxwell action
 \begin{equation}
  S = \frac{1}{16\pi G} \int d^4 x \sqrt{-g} \left( R - F^{\mu\nu} F_{\mu\nu} \right)
\end{equation} 
 where $F^{\mu\nu}$ is  the Maxwell field strength \cite{30}.  In terms of the electromagnetic potential
 ${\cal A}_{\mu}$ and the Dunkl derivatives, it reads as
 \begin{equation}
  F_{\mu\nu} = D_{\mu} {\cal A}_{\nu} - D_{\nu} {\cal A}_{\mu}.\label{1} 
\end{equation}  
In the present study,  the following  electromagnetic potential $A_{\mu}$  form  
 \begin{equation}
  {\cal A}_\mu = ( \frac{Q}{4\pi\varepsilon_0 r}, 0, 0, 0)
\end{equation}      
 is considered.  Using gauge theory tools,   the  corresponding  energy-momentum tensor is found to be  \begin{equation}
  T_{\mu\nu} = \frac{1}{4\pi} \left( F_{\rho\mu} F^{\rho\beta} g_{\nu\beta} - \frac{1}{4} g_{\mu\nu} F_{\rho\beta} F^{\rho\beta} \right).\label{2}
\end{equation} 
With the help of  Eq.(\ref{1}),  one can calculate the  components of the tensor $F_{\mu\nu}$.  The computations lead to 
 \begin{equation}
 \begin{aligned}
F_{rt} =-F_{tr} &= D_r {\cal A}_t(r) = -\frac{Q}{4\pi\varepsilon_0 r^2}\left(1-\sum_{i=1}^{3}\alpha_{i}(1-\mathcal{R}_{i})\right)^{\!2}.
\end{aligned}
 \end{equation}
Similarly, the  contra-variant components of $
 F^{\mu\nu} = g^{\mu\alpha} g^{\nu\beta} F_{\alpha\beta}$ can be expressed as follows 
\begin{equation}
F^{rt} = -F^{tr} = \frac{Q}{4\pi\varepsilon_0 r^2}\left(1-\sum_{i=1}^{3}\alpha_{i}(1-\mathcal{R}_{i})\right)^{\!2}.
\end{equation}
Using  Eq (\ref{2}),  we can obtain  the non-zero components of $T_{\mu \nu}$.  They are shown to be expressed as 
 \begin{equation}
  \begin{aligned}
T_{tt} &= \frac{Q^2 f}{32\pi^3\varepsilon_0^2r^4}\left(1-\sum_{i=1}^{3}\alpha_{i}(1-\mathcal{R}_{i})\right)^{\!2} \\
T_{rr} &= \frac{-Q^2}{32\pi^3\varepsilon_0^2r^4 f}\left(1-\sum_{i=1}^{3}\alpha_{i}(1-\mathcal{R}_{i})\right)^{\!2} \\
T_{\theta\theta} &= \frac{Q^2}{32\pi^3\varepsilon_0^2r^2}\left(1-\sum_{i=1}^{3}\alpha_{i}(1-\mathcal{R}_{i})\right)^{\!2} \\
T_{\phi\phi} &= \sin^2\theta\, T_{\theta\theta}.
\end{aligned}
 \end{equation}
By inserting the curvature components and the energy-momentum tensor into the  Einstein field equations
\begin {equation} G_{\mu\nu} = R_{\mu\nu} - \frac{1}{2} R g_{\mu\nu} = 8\pi G T_{\mu\nu}, \end{equation}  and  taking   $\dfrac{G}{4 \pi^{2} \varepsilon^{2}_{0}}=1$, we extract a couple of differential equations  satisfied by  the  black hole metric function \( f(r) \)
\begin{align}
-\frac{f'}{2r}(2 + \dfrac{\zeta}{2}) - \dfrac{3}{2}\frac{f}{r^2}\zeta- \frac{5}{16}\dfrac{f}{r^2}\zeta^2 - \frac{f}{r^2} + \frac{1 + \delta}{r^2} &= \frac{Q^2}{r^4}(1 - \dfrac{\zeta}{2})^2, \label{11}\\
-\frac{f' r}{2}(2 + \dfrac{3}{2}\zeta) +  \dfrac{f}{2} \zeta \left(1 + \frac{5}{8}\zeta \right) + \frac{r^2 f''}{2} &= \frac{Q^2}{r^2}(1 - \dfrac{\zeta}{2})^2,\label{22} 
\end{align}
where  $\zeta$ is a new Dunkl parameter   given by 
\begin{equation}
\zeta= 2\sum_{i=1}^3 \alpha_i(1 - \mathcal{R}_i).
\end{equation}
Adding  Eqs.(\ref{11}) and (\ref{22}) gives the following unified expression
\begin{equation} r^2 f'' + r \zeta f'- 2f (1 + \zeta)   + 2(1 + \delta) = \frac{4Q^2}{r^4}(1 - \dfrac{\zeta}{2})^2. 
\end{equation}
 Solving this system yields the charged deformed metric function
\begin{equation}
f(r) = \frac{1 + \delta}{1 + \zeta} + \frac{c_{1}}{r^{1 + \zeta} }+ c_{2} r^2 + \frac{Q^2}{r^2} \left(1 - \frac{\zeta^2}{4(1 + \zeta)} \right),\label{o}
\end{equation}
where \( c_{1} \) and \( c_{2} \) are   dimensioned constants of  integration.  Their dimensions are 
\begin{equation}
 [c_{1}]= [L]^{1+\zeta}, \qquad   [c_2]= [L]^{-2}
\end{equation}
where $[L]$  is the fundamental dimension of the  length.  Before  approaching  the optical behaviors of such a solution, we provide  a couple of comments on the obtained solution:
\begin{itemize}
\item This extended charged metric function can recover certain known solutions. Turning off  the  Dunkl deformations  where one has    \( \zeta = \delta= 0 \), the metric  function \( f(r) \) reduces to
\begin{equation}
f(r) = 1 + \frac{c_1}{r} +c_{2} r^2 + \frac{Q^2}{r^2}.
\end{equation}
Taking  the  normalized units  and  the following  identification
 \begin{equation}\label{am}
c_1= -2M, \qquad c_2= -\frac{\Lambda}{3},
	\end{equation}
 we get 
 \begin{equation}\label{am}
	f(r) =1-\frac{2M}{r} -\frac{\Lambda}{3}r^2+   \frac{Q^2}{r^2},
	\end{equation} 
representing  the well-known Reissner–Nordström–de Sitter solution where $M$ and   $\Lambda$  are interpreted as  the mass parameter  and   the cosmological constant, respectively \cite{30}. 
\item  The limit $Q=0$  does not recover  directly   the metric function obtained in  \cite{9} given by 
\begin{equation}
f_{\xi}(r)=\frac{1}{(1+\xi)}-2 M r^{\frac{1}{2}(1-\sqrt{9+8 \xi})}-\frac{\Lambda}{3} r^{\frac{1}{2}(1+\sqrt{9+8 \xi})}, \label{d}
\end{equation}
where  $\xi$  is a parameter expressed in terms of  the Dunkl reflections. We  believe  that  the electric charge $Q$ has   brought  a significant contribution to the deformation structure of the   obtained charged solutions. However,  a comparison could be  elaborated  by  considering  $\Lambda=0$ and $Q=0$ for the two solutions given by
\begin{equation}
f_{\xi}(r)=\frac{1}{(1+\xi)}-2 M r^{\frac{1}{2}(1-\sqrt{9+8 \xi})}, \qquad 
f(r) =  \frac{1 + \delta}{1 + \zeta}  +c_1 r^{-(1 + \zeta)}.
\end{equation} 
Forgetting about the physical  dimension and  performing a limiting expansion of the functions
$\frac{1}{(1+\xi)}$ and $\frac{1}{2}(1-\sqrt{9+8 \xi})$, 
we recover the result  obtained in \cite{9} by taking 
\begin{equation}
\zeta=\dfrac{\xi}{6}, \qquad 
\delta=-\dfrac{\xi}{3}(2\xi+1).
\end{equation}
\item  The second term $\frac{c_{1}}{r^{1 + \zeta}}$ could find a place in the study of   the Newton constant  $G$   variation explored in  finding  an alternative to the
dark matter \cite{32}.    Recently,  the variation of   $G$ and  certain corrections to the newtonian gravitational
 potential have been investigated in \cite{3200}.  A close inspection shows that   the   integration constant $ c_{1}$ term   could provide a  radial expansion of   $G$.  Assuming that the
 Dunkl parameter  $\zeta$ is extremely small,  the gravitational like-potential  $\frac{c_{1}}{r^{1 + \zeta}}$  can  be expanded as
\begin{equation}
\frac{c_{1}}{r^{1 + \zeta}} = %\frac{c_{1}}{r} \sum\limits_{\ell=0}^{+\infty} \frac{(-1)^\ell}{\ell!}\zeta^\ell \ln(r)^\ell=
 \frac{c_{1}}{r}\left(1-\zeta \ln(r)+ \frac{1}{2}\zeta^2 (\ln(r))^2- \frac{1}{6}\zeta^3 (\ln(r))^3+\ldots\ldots \right)
\end{equation}
 where   the $  1/r$ term signifies the standard inverse  term and the remaining ones denote the corrected contributions.
  We believe  that  a possible link with such activities could be elaborated by  implementing the Dunkl deformations  in the spacetime geometry.   As far as we can see that   this is  a tough  task  which needs advanced thinkings.   It is pointed out, though, that we in principle could consider such  complicated  scenarios. However,    we will restrict
ourselves to  consider  constant scenarios  as they allow us to extract the corresponding   shadow  optical  behaviors  in a
straightforward manner.    We leave the other issues  for future works.
 
\end{itemize}

Working out the  $f(r)$ moduli space   computations   amounts to approaching 
the  thermodynamical and the optical properties. This is a highly non-trivial task as it  may require
handling complicated  algebraic  equations. The general  discussion  is beyond the scope of
the present work, though we will  consider the  analysis  in the case of the following identification \begin{equation}\label{am}
c_1= -2M \epsilon_M\, \qquad c_2= 0,  %\qquad  c_1= -2M^{1+\zeta}, \qquad c_2= 0
	\end{equation}
where  $\epsilon_M$ is a free  constant with dimension  $[L]^{\zeta}$.   In this way,   the  treated fundamental metric function $f(r)$   takes the following form 
  \begin{equation} 
f(r)=\frac{1+\delta }{\zeta +1}-\frac{2 M\epsilon_M}{r^{\zeta +1}}-\frac{\left(\zeta ^2-2 \zeta -2\right) Q^2}{(2 (\zeta +1)) r^2}.
\label{metr}
\end{equation}
It has been shown    that the metric function encodes the physical behaviors of the associated black holes in terms of the involved parameters.    It has been  remarked that there are of course important differences and further similarities between the present solutions and the previous ones.

To investigate the physical  properties of the obtained charged Dunkl solutions, we first analyze the behavior of the black hole metric function.  Precisely,   one should identify  regions of the moduli space in which the black hole solution admits at least one real event horizon radius.   This can be described as a very complex  piece of work, as it requires greater technical efforts. Through numerical simulations,  it is possible to handle and examine these solutions. However,  explicit analytical solutions   can be obtained by fixing the value of the parameter $\zeta$ and solving the horizon equation $f(r)=0$.  Indeed, they can  be derived at least  for the two special cases $\zeta = 0$ and $\zeta = 1$. These are given by
\begin{equation}
r_h = 
\begin{cases} 
\dfrac{M \pm \sqrt{(M\epsilon_M)^{2} - (\delta+1)Q^{2}}}{\delta+1} & \text{for } \zeta = 0,\\[2mm]
\pm \dfrac{\sqrt{8M\epsilon_M - 3Q^{2}}}{\sqrt{2(\delta+1)}}, & \text{for } \zeta = 1.
\end{cases}
\end{equation}

\begin{figure}[!ht]
\begin{center}
\centering
\begin{tabbing}
\centering
\hspace{8.cm}\=\kill
\includegraphics[width=8cm, height=7cm]{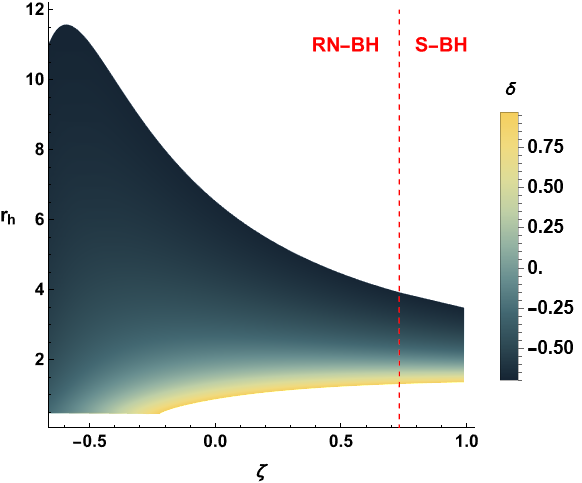}
\hspace{0.1cm} \includegraphics[width=8cm, height=7cm]{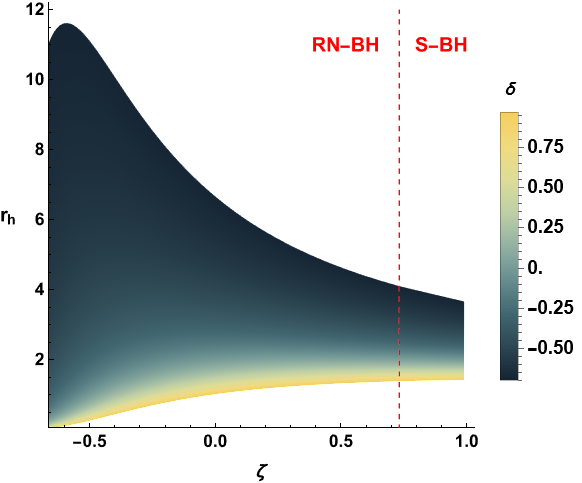}\\
  \end{tabbing}
\caption{  \it \footnotesize Left: Black hole horizons in terms of $\zeta$ and $\delta$ for $Q=0.5$ . Right:  Black hole horizons in terms of $\zeta$ and $\delta$ for $Q=0.01$.}
\label{F1}
\end{center}
\end{figure}

 A careful examination shows that an analytical solution for the constraint $f(r)=0$ for generic values of the parameters $\lbrace\zeta, \delta \rbrace$ is not ensured. It is therefore necessary to develop a numerical approach to determine  possible solutions. To do so, we employ a parallel computing program with CUDA to exploit the GPU architecture \cite{3200,3201}. This helps to accelerate the used  numerical method, providing an effective approach to solve the   event horizon constraint  $f(r)=0$. Accordingly, the obtained solutions for acceptable ranges  for the parameters  $(\zeta, \delta)$ are presented in Fig.(\ref{F1}) by  considering  $  M\epsilon_M =1$. 
It follows that  the  black hole horizons seem to have a maximal  value.  It has been observed that the parameter  $\zeta$ separates  the black hole horizons in  two types,  the Reissner-Nordström (RN)  solutions for $\zeta < 0.732$ and  the Shwarzschild (S)  solutions for $\zeta \geq 0.732$. Moreover,  the black hole horizon radius   decreases by increasing the parameter  $\delta$.  The region under the yellow teal corresponds to naked singularities for  RN  solutions and a minimal black hole size for Shwarzschild solutions.  It has been observed that  the decrease in charge reduces the values of $\zeta$ for which the solution  behaves like  a naked singularity.

\begin{figure}[h!]
\begin{center}
	\includegraphics[scale=1]{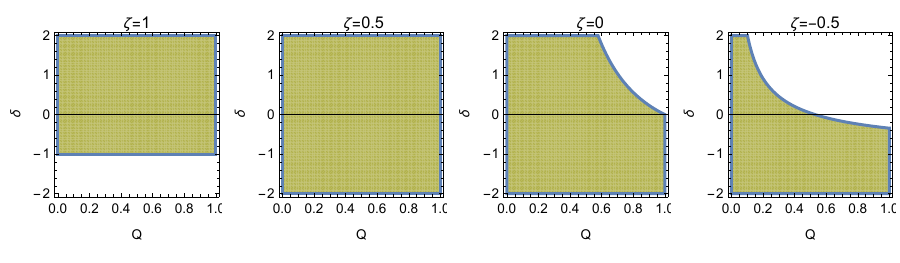}
\end{center}

\caption{ \it \footnotesize Parameter region of the moduli space ensuring the existence of a real event horizon for the charged Dunkl black hole by taking  $  M\epsilon_M =1$.}
\label{H1}
\end{figure}

In the present study, however, we examine  the effect of the parameter $\delta$ for fixed values of $\zeta$.   In Fig.(\ref{H1}), we depict the acceptable regions of the  reduced moduli space. 
 As  this figure  shows,  the negative values of $\zeta$ strongly reduce the region of $\delta$ for which an event horizon exists. For $\zeta = 1$, only values between $-1$ and $2$ are considered in the subsequent analysis. Interestingly, the case $\zeta = 0.5$ produces a fully squared region, which enlarges the moduli space parameter range compared to the case $\zeta = 0$. In the latter case, there is also a restriction on the values of the charge  $Q$, which must lie between $0.6$ and $1$.  Taking $Q = 0.7$ and $\zeta = 0.2$, for different allowed values of $\delta$, Fig.~(\ref{2}) illustrates the behavior of the metric function. Concretely, there exists  two real solutions for each considered value of 
$\delta$, indicating a non-extremal black hole structure.

\begin{figure}[h!]
\begin{center}
	\includegraphics[scale=1]{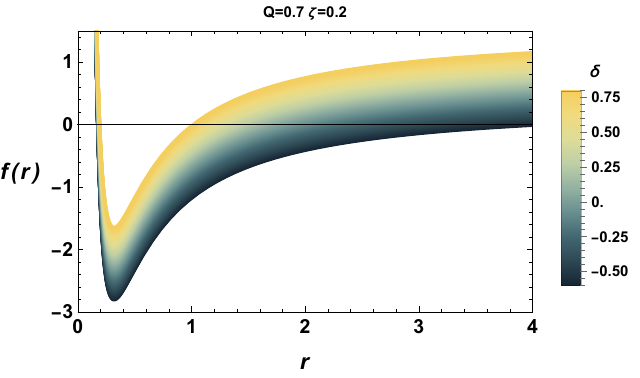}
\end{center}
\caption{  \it \footnotesize Metric function behaviors for $M\epsilon_M=1$, $Q=0.2$,  and $\zeta = 0.5$.}
\label{2}
\end{figure}
\begin{figure}[h!]
\begin{center}
	\includegraphics[scale=1]{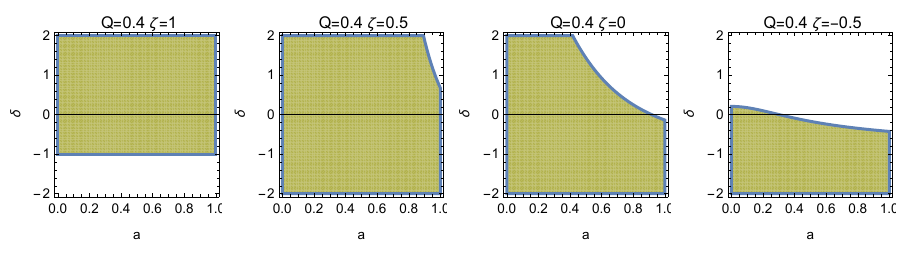}
\end{center}
\caption{Regions in the $(\delta, a)$--plane, 
where the metric admits at least one real event horizon radius for  $M\epsilon_M=1$.}
\label{3}
\end{figure}

Having discussed  the non-rotating  solution behaviors, we now turn to the building  of rotating  and charged Dunkl black holes. To obtain such solutions, we employ the Newman-Janis algorithm without complexification \cite{33}.   This  allows   the metric to be expressed as in \cite{34,35}.  It is denoted that this  method can be extended to  a variety of modified gravity models by introducing additional parameters originated either from the spacetime geometry or the  implemented matter  physical fields including the dark energy.   Based on such an algorithm, we could  investigate the physical properties of  the rotating version of the  charged   Dunkl black holes.  Adopting the Boyer-Lindquist coordinate system,  we  can get the following line element for the  black hole metric
\begin{equation}
ds^2=\left( \frac{\sigma(r)}{\Sigma(r)} -1\right) dt^2-\frac{2a^2 \sigma(r)}{\Sigma(r)}\sin^{2} \theta dtd\phi+ \left( r^2+a^2+\frac{a^2 \sigma(r) \sin^{2} \theta }{\Sigma(r)}\right)\sin^{2} \theta d\phi^2 +\frac{\Sigma(r)}{\Delta(r)}dr^2+\Sigma(r)d\theta^2,
\label{mr}
\end{equation}
where one has used
\begin{eqnarray}
\Sigma(r)=r^2+a^2 \cos^2\theta,\quad
  \Delta(r)= f(r)r^2+a^2,\quad 
 \sigma(r) = r^2- r^2f(r).  
\end{eqnarray}
In this solution,   $a$  denotes  the rotation parameter of the charged  Dunkl  black holes. This rotating metric is derived under conditions that ensure it provides a physically acceptable solution of the field equations, with the energy-momentum tensor interpreted as an imperfect fluid rotating about the $z$-axis. To illustrate the influence of the rotation parameter on the metric function, 
 Fig.~(\ref{3}) displays the regions in the $(\delta, a)$--plane, for a fixed value of 
the parameter $\zeta$, where at least one real event horizon radius exists. As shown in the figure,  the negative values of $\zeta$ significantly reduce the 
allowed ranges of the  moduli space parameters in the rotating case. 
For $\zeta = 1$, the rotation parameter has no effects compared to the 
non-rotating case, while for other values it slightly decreases the size of the 
allowed region. In the following section, we study the optical  properties of such black 
holes for parameter values lying within the allowed regions.

\section{Shadow  behaviors of rotating and charged Dunkl black holes}
The main objective of this section is to investigate the optical properties of  non-rotating and  rotating  charged black holes
by analyzing their shadow structures. To this end, 
we employ the Hamilton--Jacobi formalism via  the Jacobi action relation
\begin{equation}
\frac{\partial \mathcal{S}}{\partial \sigma} 
= -\frac{1}{2} g^{\mu \nu} 
\frac{\partial \mathcal{S}}{\partial x^\mu} 
\frac{\partial \mathcal{S}}{\partial x^\nu}.
\end{equation}
According to \cite{36}, this action  can be separated as
\begin{equation}
\mathcal{S} = -Et + L\phi + S_r(r) + S_\theta(\theta)
\end{equation}
where $E$ denotes the conserved energy, $L$ is  the conserved angular momentum, 
$S_r(r)$ denotes a function of the radial coordinate $r$. $S_\theta(\theta)$  is 
a function of the polar angle $\theta$.
The corresponding null geodesic equations for the radial coordinate $r$  and the angular one    $\theta$
can be obtained using  the variable separation  following  a manner analogous to the Carter 
mechanism \cite{37}. For non-rotating charged Dunkl black holes,  we  can find
\begin{equation}
\begin{aligned}
r^2(r)\left(\frac{dS_r(r)}{dr}\right)^2 + L^2 - \frac{r^2}{f(r)}E^2 &= -\mathcal{C}, \\
\left(\frac{dS_\theta(\theta)}{d\theta}\right)^2 + L^2 \cot^2\theta &= \mathcal{C},
\end{aligned}
\label{se}
\end{equation}
where $\mathcal{C}$ is a separation constant. This yields to the following  equations of motion 
\begin{align}
\dot{t} &= -\frac{E}{f(r)}, \nonumber \\
\dot{r} &= \pm \sqrt{\mathcal{R}(r)},  \nonumber\\
r^2\dot{\theta} &= \pm \sqrt{\mathcal{C}-L^2 \cot^2\theta}, \\
\dot{\phi} &= \frac{L}{h(r)\sin^2\theta} \nonumber
\end{align}
where  $\mathcal{R}(r)$ is a radial function taking  the form
\begin{equation}
\mathcal{R}(r) = E^2 \left( 1 - \frac{f(r)}{r^2}(\Xi ^2+\eta)\right).
\end{equation}
In this function,   $\eta$ and $\Xi$  represent  the dimensionless impact parameters expressed as follows
\begin{equation}
{{\eta}} = \frac{\mathcal{C}}{E^2}, \qquad  { \Xi} = \frac{L}{E}.
\end{equation}
The shadow boundary is determined by the conditions for unstable circular photon 
orbits, namely
\begin{equation}
\mathcal{R}(r)\big|_{r=r_o} = 0, \qquad 
\frac{\mathcal{R}(r)}{dr}\big|_{r=r_o} = 0,
\label{sheq}
\end{equation}
where $r_o$ denotes the radius of the circular photon orbit. Solving this system requires fixing the value of the parameter $\zeta$  in order to get  such a radius. For instance, two explicit solutions can be obtained   by taking 
\begin{equation}
r_0 =
\begin{cases}
\dfrac{\sqrt{8M\epsilon_M - 3Q^2}}{\sqrt{1+\delta}}, & \zeta = 1, \\[1.2em]
\dfrac{3M + \sqrt{9(M\epsilon_M)^2 - 8Q^2 - 8Q^2\delta}}{2(1+\delta)}, & \zeta = 0.
\end{cases}
\end{equation}
For   $ \zeta = \delta=Q= 0$ and $\epsilon_M=1$, we recover  the value 
$r_0 =3M$ associated with the  ordinary  Schwarzschild  black hole \cite{39}. The above impact parameters can be obtained by solving  Eq.(\ref{sheq}). The computation provide
\begin{equation}
\Xi ^2+\eta=\frac{2 (\zeta +1) r_0^{\zeta+4}}{r_0^{\zeta} \left(\left(-\zeta ^2+2 \zeta +2\right) Q^2+2 (\delta +1) r_0^2\right)-4 M\epsilon_M (\zeta +1) r_0}.
\end{equation}
The apparent shape of the black hole shadow, as observed at spatial infinity, 
can be characterized by the celestial coordinates $(X,Y)$ being 
\begin{equation}
\begin{aligned}
X &= \lim_{r_{ob}\to\infty}\left(-r_{ob}^2 \sin\theta_{ob}\,\frac{d\phi}{dr}\right), \\
Y  &= \lim_{r_{ob}\to\infty}\left(r_{ob}^2 \,\frac{d\theta}{dr}\right),
\end{aligned}
\end{equation}
where $r_{ob}$ is the distance of the observer from the black hole and 
$\theta_{ob}$ is the inclination angle between the observer line of sight and 
the symmetry axis of the black hole. The coordinate $X$ corresponds to the 
apparent perpendicular displacement from the axis of symmetry, while $Y$ 
measures the displacement from the projection onto the equatorial plane. For 
null geodesics, the shadow boundary satisfies
\begin{equation}
X^2 + Y^2 = \Xi^2 + \eta.
\end{equation}

\begin{figure}[t] %[!ht]
		\begin{center}
		\centering
			\begin{tabbing}
			\centering
			\hspace{7.5cm}\=\kill
			\includegraphics[scale=.7]{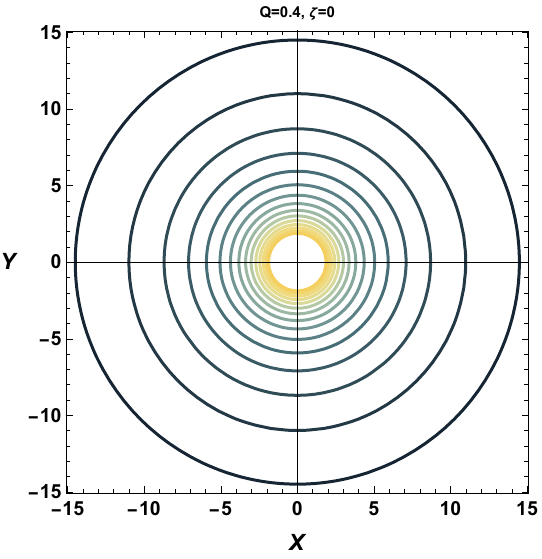} \>
			\includegraphics[scale=.7]{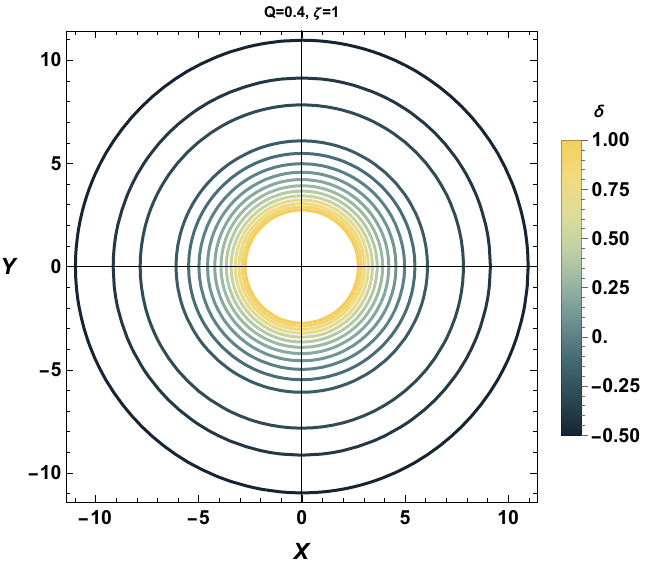} \\
		   \end{tabbing}
\caption{{\it \footnotesize  Non-rotating shadow behaviors in  the celestial coordinates $(x,y)$     by varying $\delta$ and $\zeta$ and taking  $M\epsilon_M=1$.}}
\label{SF1}
		   \end{center}
\end{figure}

In Fig.~(\ref{SF1}), we present the shadow profiles for $\zeta=0$ and $\zeta=1$ while keeping the mass and the  charge fixed and varying the parameter $\delta$. The results indicate that increasing the value of $\zeta$ leads to a reduction in the overall shadow size without introducing any noticeable deformation. On the other hand, increasing $\delta$ from negative to positive values also reduces the shadow size. For both Dunkl parameters, no deformation effects  have been  observed. However, the influence of $\delta$ becomes more pronounced as $\zeta$ increases. This demonstrates that the combined effect of these parameters primarily controls the size of the shadows. Now,  we  examine the effect of the rotation parameter on such black   solutions by studying how the parameter  space determines both the size and  the shape of the resulting shadows.
Starting from the metric (\ref{mr}) and employing the Hamilton–Jacobi separation method, we derive the following set of four equations of motion
\begin{align}
\Sigma \dot{t}& = \frac{r^2+a^2}{\Delta}\left[ E\left( r^2+a^2\right) -a L\right] 
+a\left[  L-aE\sin^2\theta \right]  \\
(\Sigma \dot{r})^2 &=\mathcal{R}(r) \\
( \Sigma\dot{ \theta})^2 & =\Theta(\theta)\\
\Sigma \dot{\phi} &= \left[ L \csc ^2 \theta-aE\right] 
+\frac{a}{\Delta} \left[ E\left(r^2+a^2\right) -aL\right],
\end{align}
where $ \mathcal{R}(r)$  and $\Theta(\theta) $  are radial functions  given by
 \begin{align}
\mathcal{R}(r) &= \left[ E \left( r^2 + a^2 \right) - a L \right]^2 
- \Delta \left[ \mathcal{C} + \left( L - a E \right)^2 \right], \\
\Theta(\theta) &= \mathcal{C} - \left( L \csc \theta - a E \sin \theta \right)^2 
+ \left( L - a E \right)^2.
\end{align}
 Considering the unstable spherical photon orbits and projecting the null trajectories onto the observer celestial plane, the apparent shadow boundary is governed by the two impact parameters 
 \begin{align}
\eta 
&= \frac{r^{2}\!\left[\,16 a^{2}\Delta(r) + 8 r\,\Delta(r)\Delta'(r) - 16 \Delta(r)^{2} - r^{2}\Delta'(r)^{2}\right]}{a^{2}\,\Delta'(r)^{2}}\bigg|_{r=r_0}, \\[4pt]
\Xi 
&= \frac{(r^{2}+a^{2})\,\Delta'(r) - 4 r\,\Delta(r)}{a\,\Delta'(r)}\bigg|_{r=r_0}.
\end{align}

 \begin{figure}[!ht]
		\begin{center}
		\centering
			\begin{tabbing}
			\centering
			\hspace{2.cm}\=\kill
			\includegraphics[scale=0.55]{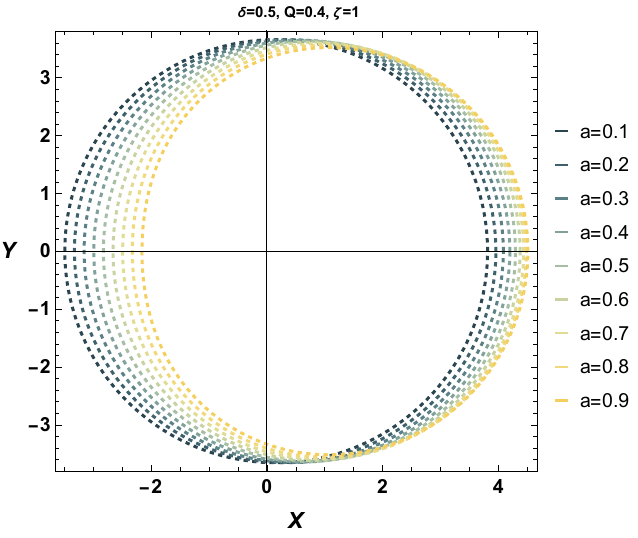} 
	\hspace{0.1cm}		\includegraphics[scale=0.5]{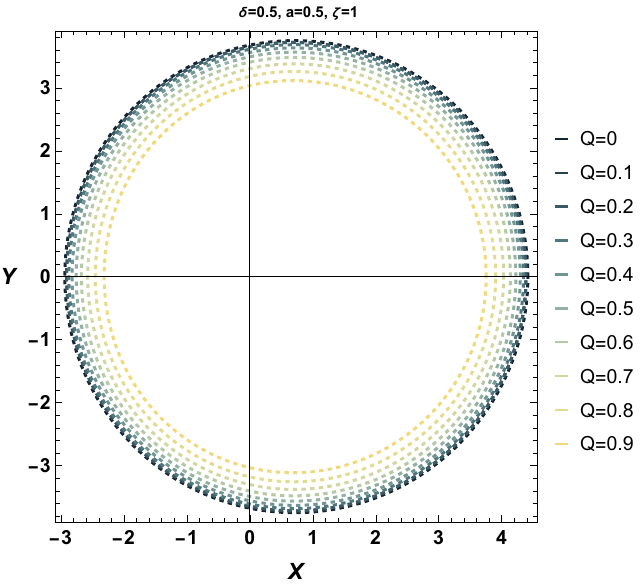}\hspace{0.1cm}	\includegraphics[scale=0.5]{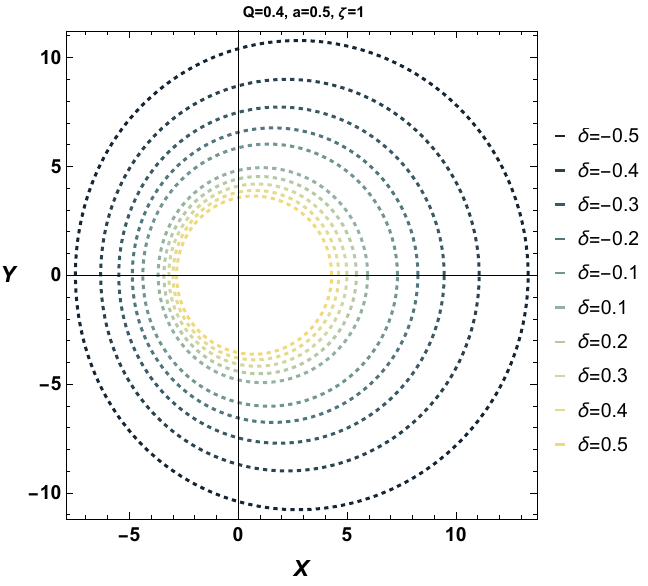}\\ 
	
		   \end{tabbing}
\caption{ \it \footnotesize Behavior of the black hole shadow as the charge, the rotation, and the Dunkl parameter $\delta$ are varied with $M\epsilon_M=1$.}
\label{rsa}
\end{center}

\end{figure}
In  Fig.~(\ref{rsa}), we  illustrate the black hole shadow behaviors in terms of the moduli space parameter. This reveals  the effects of the charge, the rotation parameter, and the  Dunkl  parameter $\delta$. We do not take into consideration variations in the parameter $\zeta$, since its only effect is to amplify that of $\delta$.  Hence, we focus on the variation of $\delta$ for a fixed $\zeta$.  The rotation parameter maintains its effect by deforming the shadow into a D-shape, while the charge primarily acts to slightly reduce its overall size. On the other hand, increasing the  parameter $\delta$ from $-0.5$ to $0.5$ significantly decreases the size of the shadow without altering its shape.
 
 Taking into account the actual event horizon radius and the observed shadow radius impose constraints on the model parameters. In particular, choosing a positive $\zeta$ and $\delta > -0.5$  appears physically reasonable.  However,  to  ensure that these theoretical predictions are fully consistent with reality, a careful comparison with observational data, such as measurements of M$87^*$ or Sgr $A^*$, is required.

\section{Energy emission rate}

\begin{figure}[!ht]
		\begin{center}
		\centering
			\begin{tabbing}
			\centering
			\hspace{2.cm}\=\kill
			\includegraphics[scale=0.52]{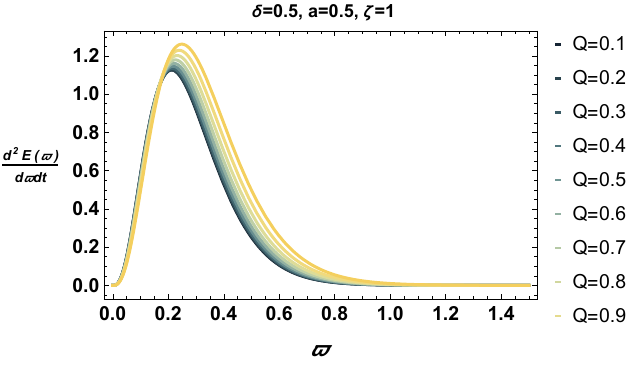} 
	\hspace{0.1cm}		\includegraphics[scale=0.52]{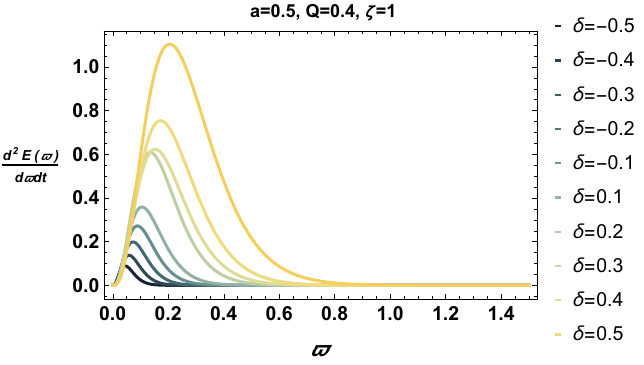}\hspace{0.1cm}	\includegraphics[scale=0.52]{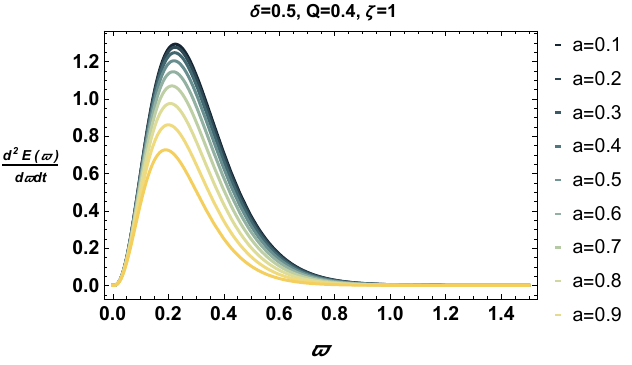}\\ 
	
		   \end{tabbing}
\caption{ \it \footnotesize Energy emission rate versus emission frequency for different values of the charge, the rotation, and the Dunkl parameter $\delta$ with $M\epsilon_M=1$.}
\label{EER}
\end{center}

\end{figure}

In this section, we turn our attention to the energy emission rate associated with  rotating and  charged 
 black holes in the  deformed  Dunkl spacetime. For a distant observer, the absorption cross section at very high energies 
asymptotically approaches its geometrical optics limit, which is directly related to 
the black hole shadow. In intermediate regimes, the absorption cross section exhibits 
oscillations around a constant limiting value, denoted by $\sigma_{\text{lim}}$. This constant 
has been shown to coincide with the geometrical cross section of the photon sphere, 
as determined by the properties of null geodesics \cite{40,41,42}. Since the shadow 
encodes the optical appearance of the black hole, it can be identified with this limiting 
value, allowing one to approximate  $\sigma_{\text{lim}}$ as follows 
\begin{equation}
\sigma_{\text{lim}} \simeq \pi R_{s}^{2},
\end{equation}
where $R_{s}$ is the shadow radius. Within this framework, the differential energy 
emission rate takes the form 
\begin{equation}
\frac{d^{2}E(\omega)}{d\omega \, dt} = 
\frac{2\pi^{3} R_{s}^{2}}{e^{\omega/T_{H}}-1} \, \omega^{3},
\end{equation}
where $T_{H}$ is the Hawking temperature of the black hole and $\omega$ is the emission frequency. This relation establishes a 
clear connection between the thermodynamic properties of the black hole and its 
optical features. Indeed, this may  provide a useful tool to probe the  spacetime parameters 
through  the observational signatures. Considering the rotating metric,  the Hawking  temperature  of such black holes is given by 
\begin{equation}
T_{H}=\frac{r^{-\zeta} \left(M\epsilon_M \left(\zeta^2-1\right)+(\delta +1) r^{\zeta +1}\right)}{2 \pi  (\zeta +1) \left(a^2+r^2\right)}.
\end{equation}
In Fig.~(\ref{EER}), we illustrate the variation of the energy emission rate as a function of the 
emission frequency. The first panel corresponds to different charge values, the second panel 
shows the effect of varying the Dunkl  parameter $\delta$, while the third panel displays the 
impact of different rotation parameter values.

It has been remarked that   the parameters  $a$ and $Q$  exhibit  similar  impact  contributions contrary to $\delta$. The evaporation of the black holes increases by taking small  values  of  $a$    as expected since it decreases the black hole temperature. However,   $\delta$  increases  the energy emission rate values.

\section{Constraints on black hole parameters from EHT observations via  CUDA }

 To establish a connection between the  theoretical predictions and  the observational evidence, the following section presents an analysis of the shadow cast by rotating and charged  black holes in the  Dunkl spacetime, incorporating the observational results from the  EHT  collaboration. In particular, we utilize the EHT data for M$87^*$ and Sgr~A$^*$ to constrain the parameters of these black holes~\cite{EventHorizon,Chakhchi,Gogoi}.

The constraints can be derived using the fractional deviation from the Schwarzschild black hole shadow diameter given by 
\begin{equation}
{ d} = \frac{R_s}{r_{sh}}-1,
\end{equation}
where \( R_s \) denotes the shadow radius and \( M \) the black hole mass and $r_{sh}$ indicates the Schwarzschild one. The dimensionless quantity \( R_s/M \) serves as a key observable to compare theoretical models with empirical measurements. The 1-\(\sigma\) and 2-\(\sigma\) confidence intervals inferred from EHT observations are summarized in Table~\ref{t1}.

\begin{figure}[htbp]
\hspace*{-1.5cm}                                               
   \includegraphics[scale=0.45]{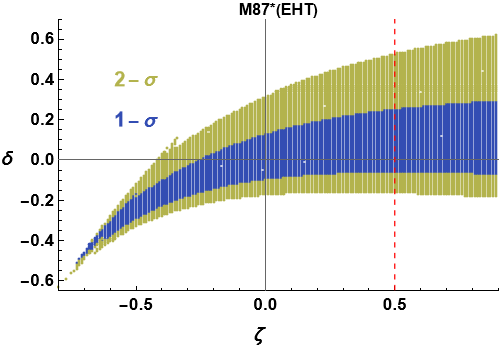}%
   \hspace{2mm}%                
    \includegraphics[scale=0.45]{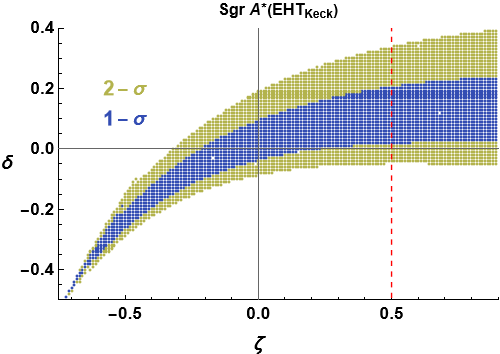}    
   \hspace{2mm}%             
    \includegraphics[scale=0.45]{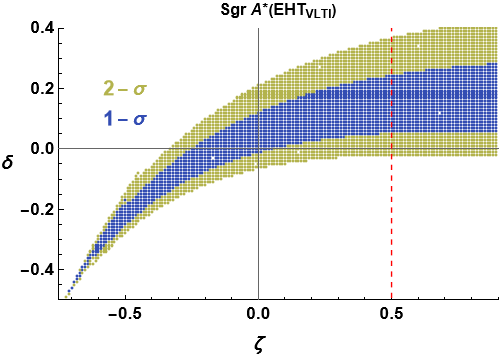}%
                           
\caption{ \it \footnotesize Constraint regions in the $( \delta,\zeta)$ plane obtained from CUDA-based simulations, showing agreement with the EHT observations of M$87^*$ and Sgr~A$^*$ within $1-\sigma$ and $2-\sigma$ confidence levels for $Q=0.4$ and  $a=0.5$ with $M\epsilon_M=1$.}

  \label{fig:Moduli}
\end{figure}

\begin{table}[h!]
\centering
\begin{tabular}{|c|c|c|c|}
\hline
\textbf{Black Hole} & \textbf{Deviation (\(d\))} & \textbf{1-\(\sigma\) Bounds} & \textbf{2-\(\sigma\) Bounds} \\ 
\hline
M87$^*$ (EHT) & $-0.01^{+0.17}_{-0.17}$ & $4.26 \leq \frac{R_s}{M} \leq 6.03$ & $3.38 \leq \frac{R_s}{M} \leq 6.91$ \\ 
\hline
Sgr~A$^*$ (EHT$_{\text{VLTI}}$) & $-0.08^{+0.09}_{-0.09}$ & $4.31 \leq \frac{R_s}{M} \leq 5.25$ & $3.85 \leq \frac{R_s}{M} \leq 5.72$ \\ 
\hline
Sgr~A$^*$ (EHT$_{\text{Keck}}$) & $-0.04^{+0.09}_{-0.10}$ & $4.47 \leq \frac{R_s}{M} \leq 5.46$ & $3.95 \leq \frac{R_s}{M} \leq 5.92$ \\ 
\hline
\end{tabular}
\caption{ \it \footnotesize Estimated fractional deviations and corresponding bounds for M87$^*$ and Sgr~A$^*$ black holes.}
\label{t1}
\end{table}
In  the present work, we employ numerical calculations based on CUDA  techniques to determine the pairs $(\zeta, \delta)$ that yield shadow configurations consistent with the experimental results within the $1-\sigma$ and $2-\sigma$ confidence intervals reported by the EHT collaboration. More specifically, for the fixed values $Q = 0.4$ and $a = 0.5$, the corresponding results are shown in Fig.(\ref{fig:Moduli}).
 
As illustrated in the three panels of this  figure, the density of points consistent with the empirical observations increases for higher values of $\zeta$ and $\delta$.  This result suggests that the spacetime background of the black hole can effectively reproduce observational signatures. Given the interdependence between the two parameters, we fix one parameter while constraining the other. For $\zeta = 0.5$, Table~(\ref{taaa}) presents the corresponding ranges of $\delta$ that yields consistency with the $1-\sigma$ and $2-\sigma$ confidence intervals for the three experimental scenarios.

\begin{table}[h!]
\centering
\begin{tabular}{|c|c|c|c|}
\hline
Confidence interval & M87$^*$ (EHT) & Sgr~A$^*$ (EHT$_{\text{VLTI}}$) & Sgr~A$^*$ (EHT$_{\text{Keck}}$) \\ 
\hline

$1-\sigma$ & $ -0.06\leq \delta\leq 0.25 $ & $ 0.06\leq \delta\leq 0.24 $& $ 0.02\leq \delta\leq 0.20 $ \\ 

\hline

$2-\sigma$ & $ -0.16\leq \delta\leq 0.52 $ & $ -0.02\leq \delta\leq 0.36 $& $ -0.04\leq \delta\leq 0.33 $ \\

\hline
\end{tabular}
\caption{ \it \footnotesize Constraints on the Dunkl  parameter $\delta$ at $\zeta = 0.5$ from EHT empirical  data.}
\label{taaa}
\end{table}
For the specific case $\zeta = 0.5$, where the metric function of the black hole exhibits a large region of parameter values leading to real and physically acceptable horizon solutions, the table indicates that only particular values of $\delta$ yield results in good agreement with the observational data. Specifically, $\delta$ should be positive and less than $0.2$, when all other parameters are fixed.

%\end{wrapfigure}

\section{Conclusions}
In this paper, we  have studied   the shadow of  a new   class of rotating and charged black holes by implementing    the  Dunkl derivative  operators.   Using such a formalism, 
we have established the  metric function encoding the physical behaviors   including the optical   aspect.   This  function has been approached  with the help of  the CUDA-accelerated simulations exploited in  machine learning activities.  Concretely, we have discussed the horizon radius behaviors. In particular, we have   determined the   regions of the module space providing physical solutions.   Applying   the Hamilton-Jacobi mechanism,  we have   elaborated  the shadow  geometrical configurations  of  non-rotating and rotating  charged  Dunkl black holes. Then, we have computed and  discussed   the energy rate of emission  for  rotating solutions. In order to enable direct comparison with current astrophysical data, we have   developed  a  CUDA high-performance numerical code to  constraint the  shadow geometries. Precisely, we have   elaborated a    numerical approach   to  impose  strict limits on the Dunkl deformation parameters   matching with   shadow observations  delivered  by the EHT collaboration.

This work paves the way for further investigations. An interesting question is to supplement this optical analysis
by addressing the angle of light deviation near these Dunkl-type black holes.  Moreover,   it  is particularly interesting to  implement  alternative contributions  to charged    Dunkl  black holes by   including  extra matter fields. We expect to be able to report on these open questions elsewhere.

\section*{Acknowledgements}
AB and HB  would like to thank N.  Askour and H. El Moumni  for collaboration on related topics. MJ   gratefully acknowledges the financial support  of the CNRST in the frame of the PhD Associate Scholarship Program PASS.

\end{document}